\pdfoutput=1
\RequirePackage{color}

\RequirePackage{ifpdf}
\documentclass{JINST}

\usepackage{graphicx}

\usepackage{multirow}

\usepackage[export]{adjustbox}
\usepackage{url}

\usepackage{color}

\usepackage[caption=false]{subfig}
\makeatletter
\AtBeginDocument{%
\DeclareRobustCommand*\subref{\@ifstar\sf@@subref\sf@subref}}
\makeatother

\usepackage{placeins}

\usepackage{datetime}

\definecolor{codebgcolor}{rgb}{0.8,0.8,0.8}
\definecolor{darkblue}{rgb}{0.0,0.0,0.3}

\DeclareGraphicsExtensions{.pdf,.png,.jpg}

\title{A Study of Dielectric Breakdown Along Insulators Surrounding Conductors in Liquid Argon}

\author{
S.~Lockwitz\thanks{Corresponding author: lockwitz@fnal.gov (S.~Lockwitz)},
and H.~Jostlein\\
Fermi National Accelerator Laboratory, P.O. Box 500, Batavia, IL, 60510, USA }

\abstract{
High voltage breakdown in liquid argon
is an important concern in the design of liquid
argon time projection chambers, which are often used as neutrino and dark matter detectors.  
We have made systematic measurements of  breakdown voltages in liquid argon along insulators surrounding negative rod electrodes where the breakdown is initiated at the anode.  The measurements were performed in an open cryostat filled with commercial grade liquid argon exposed to air, and not the ultra-pure argon required for electron drift.  While not addressing all high voltage concerns in liquid argon, these measurements have direct relevance to the design of high voltage feedthroughs especially for averting the common problem of flash-over breakdown.  The purpose of these tests is to understand the effects of materials, of breakdown path length, and of surface topology for this geometry and setup.   We have found that the only material-specific effects are those due to their permittivity.   We have found that the breakdown voltage has no dependence on the length of the exposed insulator.
A model for the breakdown mechanism is presented that can help inform future designs.

}


\begin{document}

\section{Motivation}
Liquid Argon Time-Projection Chambers (LArTPCs) are a popular detector choice for neutrino and dark matter experiments.  This technology relies upon an electric field typically on the order of 50~kV/m to drift signal electrons from ionization due to charged particle interactions in the argon.  Drift distances are typically on the scale of less than a few meters.  It has become important to understand parameters related to high voltage use in liquid argon to design future detectors and understand any technical limitations.

Dielectric breakdown in liquids, particularly transformer oils, has been actively studied for many decades,  mostly to meet the engineering needs of the power distribution industry.  Very little generic research has been done on dielectric breakdown in liquid argon.   Several recent breakdown studies~\cite{ref:bern,ref:Bay,ref:bernLatex,ref:hvc} have yielded useful results related to breakdown through the bulk liquid. This report presents a comparative study of dielectric breakdown in liquid argon along insulator surfaces surrounding high voltage conductors.  The goal is to understand these breakdowns and eventually find methods to predict and avoid them.

\section{Experiment Setup}\label{sec:setup}
\begin{figure}[hbt]
  \centering
  \subfloat[]{\includegraphics[width=0.50\textwidth,trim={5.1cm 3cm 5.4cm 3cm},clip,frame]{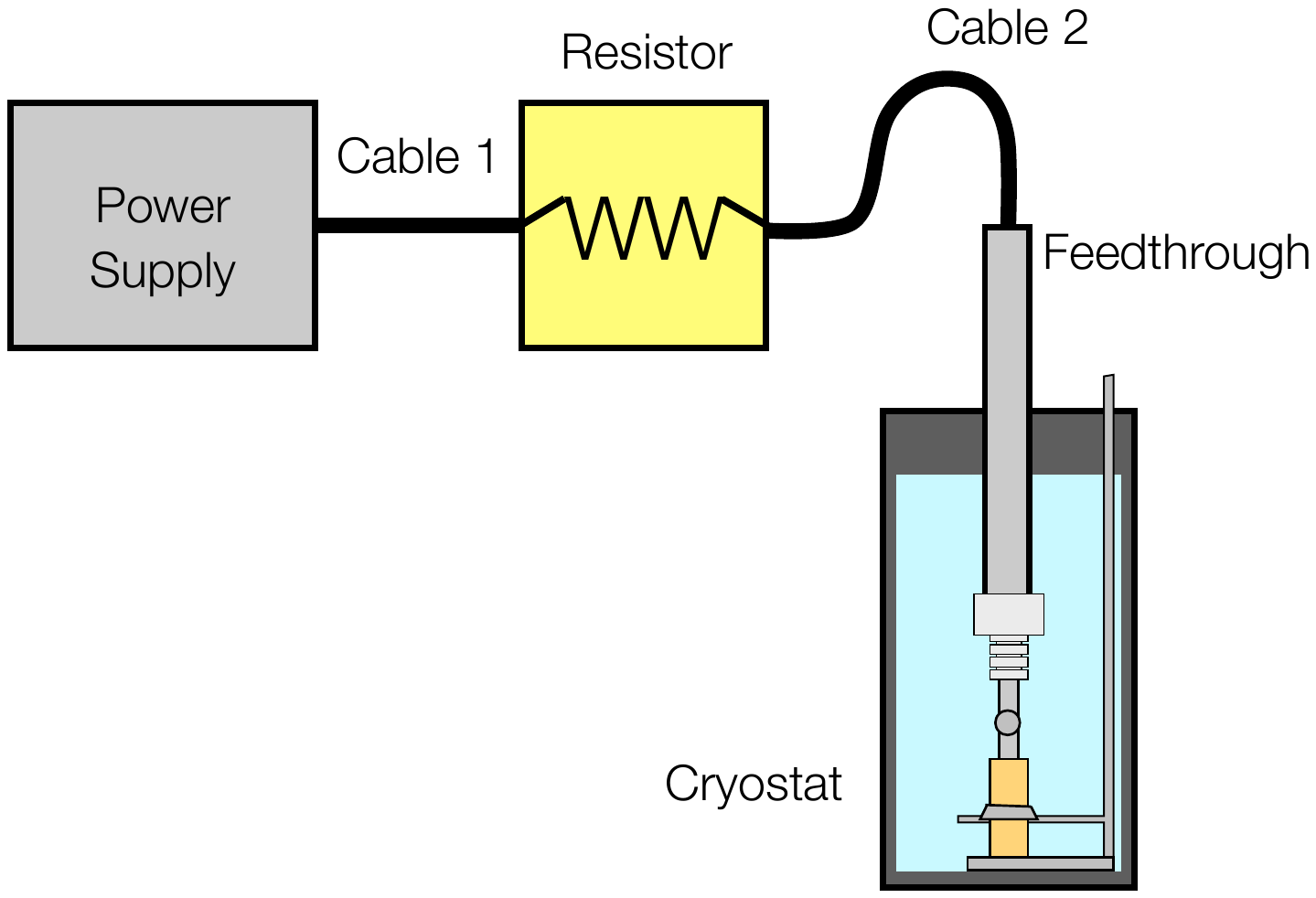}}
  \subfloat[]{\includegraphics[width=0.50\textwidth,trim={5cm 1cm 5.5cm 5cm},clip,frame]{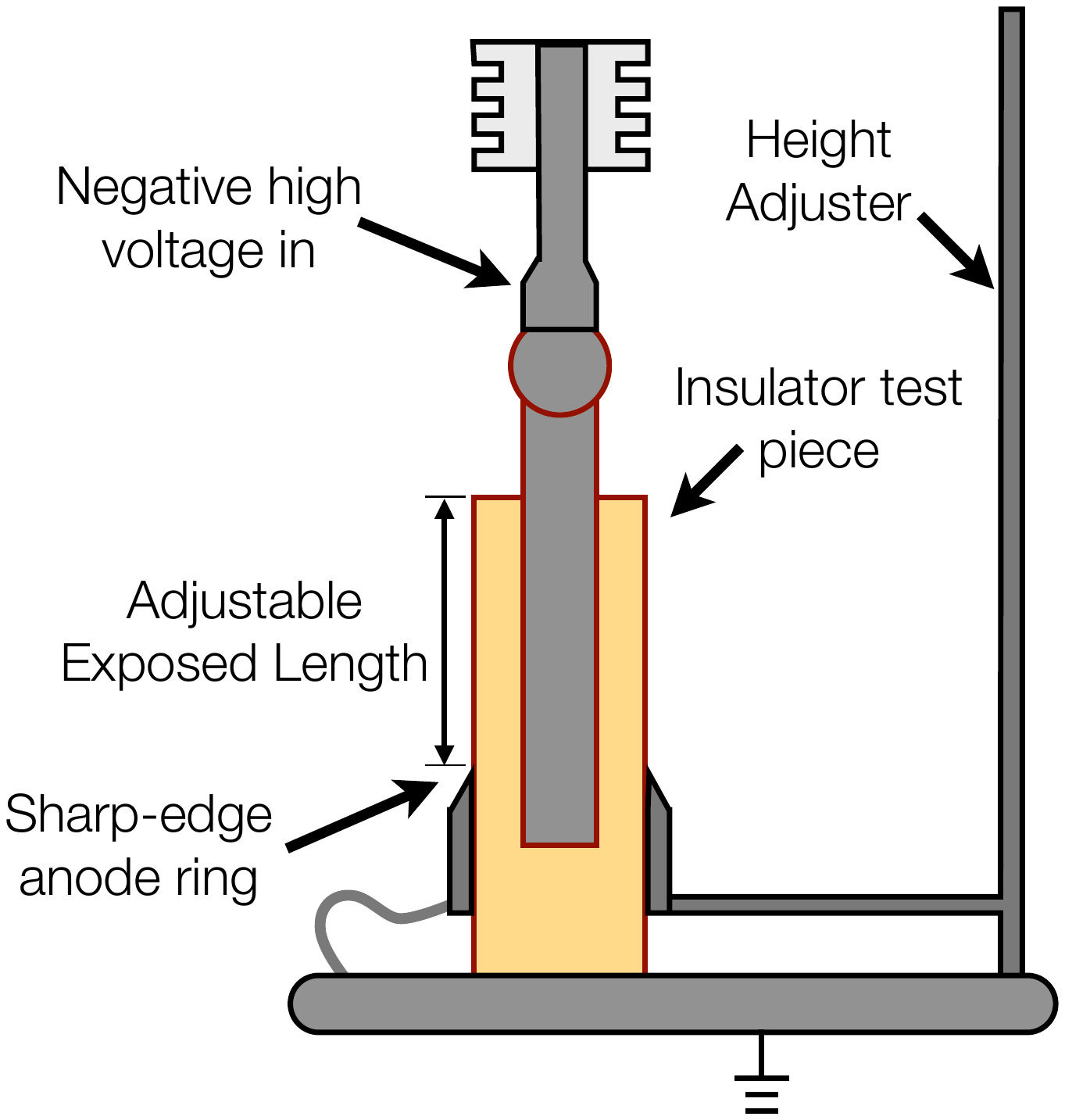}}
\caption{Drawings of the test apparatus.  The left graphic, (a), shows the complete test setup including the current-limiting resistor.  A detailed drawing including the test piece and holder is highlighted in (b).  The insulator test piece could be removed and replaced without removing the anode ring, holder and high voltage feedthrough from the liquid argon.  The high voltage was applied by a high voltage feedthrough placed on top of the test piece.  The ring could be raised and lowered from outside of the cryostat.}
\label{fig:insulatorHolderDrawing}
\end{figure}

The experiment setup is shown schematically in Figure~\ref{fig:insulatorHolderDrawing}.  We started with commercially pure liquid argon with contaminants at the parts-per-million (ppm) level.  The tests were performed in an open-to-air cryostat over the course of weeks.  This condition implies that water vapor, oxygen, and nitrogen diffused into the liquid argon over the course of the measurements. We also noted an accumulation of dust in the liquid argon, with dust especially attracted to areas of high electric field.  The effects of these contaminants are unknown.  We have not found published data on the effect of contaminants in the relevant range.

The high voltage was supplied by a Glassman LX150N12 power supply~\cite{ref:glassman} able to provide up to $-150$~kV.  The power supply was connected through an oil-insulated series resistor (75~M$\Omega$) to a high voltage feedthrough partially submersed in liquid argon.  The resistor served to partition the energy stored in the setup and limit the energy released into the cryostat during an electrical discharge.  The high voltage feedthrough was made of a stainless steel central conductor surrounded by an ultra high molecular weight polyethylene (UHMW PE) tube inside an outer stainless steel grounded tube.  The lower edge of the ground tube was encased in a UHMW PE cylinder to prevent streamer propagation along the feedthrough itself.  The lower end of the UHMW PE protruded beyond the ground tube and its surface was grooved.  The inner conductor extended further out and was machined with a concave end to accept a sphere.

Each insulator studied started in the form of a 5.1~cm diameter smooth cylinder, 30.2~cm long with a 2.54~cm diameter center-bored hole along the length of the material to a depth of 16.5~cm.  A 2.54~cm diameter stainless steel rod with a 3.8~cm diameter sphere welded on its end was inserted into each test piece as shown in Figure~\ref{fig:insulatorHolderDrawing}.  The high voltage feedthrough was placed on top of the welded sphere during testing.  Three samples were additionally tested with grooved profiles cut into the insulator surfaces.

Each piece was lowered into a grounded holder submerged in liquid argon to secure the piece during testing.  The holder had a grounded stainless steel ring attached to it with a 5.2~cm inner diameter that fit loosely around the body of the test piece.
The ring thickness flared out from the test piece at a 15 degree angle from an intentionally sharp top edge near the insulator surface to a 0.5~cm thick ring where it attached to the holder.  This top edge, the knife edge, was designed to create a high electric field to induce breakdown along the insulator surface.  The ring could be raised and lowered along the test piece by turning a threaded rod that extended outside of the cryostat.

\subsection{Experiment Procedure}\label{sec:procedure}
After insertion into the argon, each test piece was allowed to cool until there was no visible boiling before testing began.  This was done in an effort to reduce any thermal effects on spark formation.  

A {\sc{LabView}}~\cite{ref:labview} program was used to raise the high voltage in 500~V steps every 1.6~s.  When the current draw as registered by the supply was above 45~$\mu$A, the supply would trip setting the voltage to zero.  After a trip was identified, the program would wait one minute and then begin raising the voltage again.  Breakdown values were determined from a log file monitoring the voltage at sub-second intervals.

While the definition of a tripping spark is specific to our setup, it is the same for all test pieces.  There were occasions where there was an audible spark that did not trip the supply.  These events did not enter our dataset and the voltage would increase until a tripping spark occurred.

\begin{figure}[hbt]
  \centering
  \subfloat[]{\includegraphics[width=0.50\textwidth]{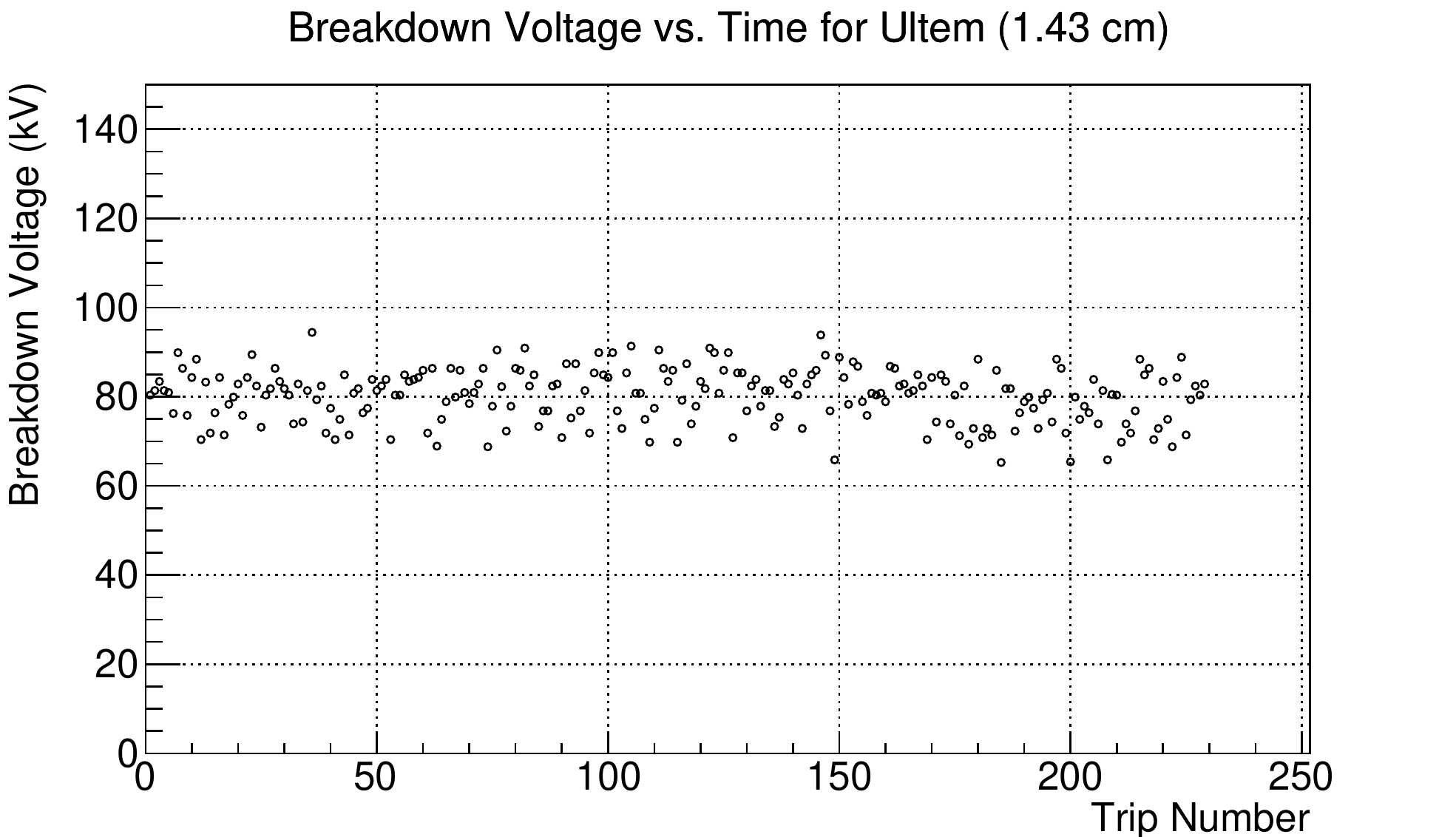}}
  \subfloat[]{\includegraphics[width=0.50\textwidth]{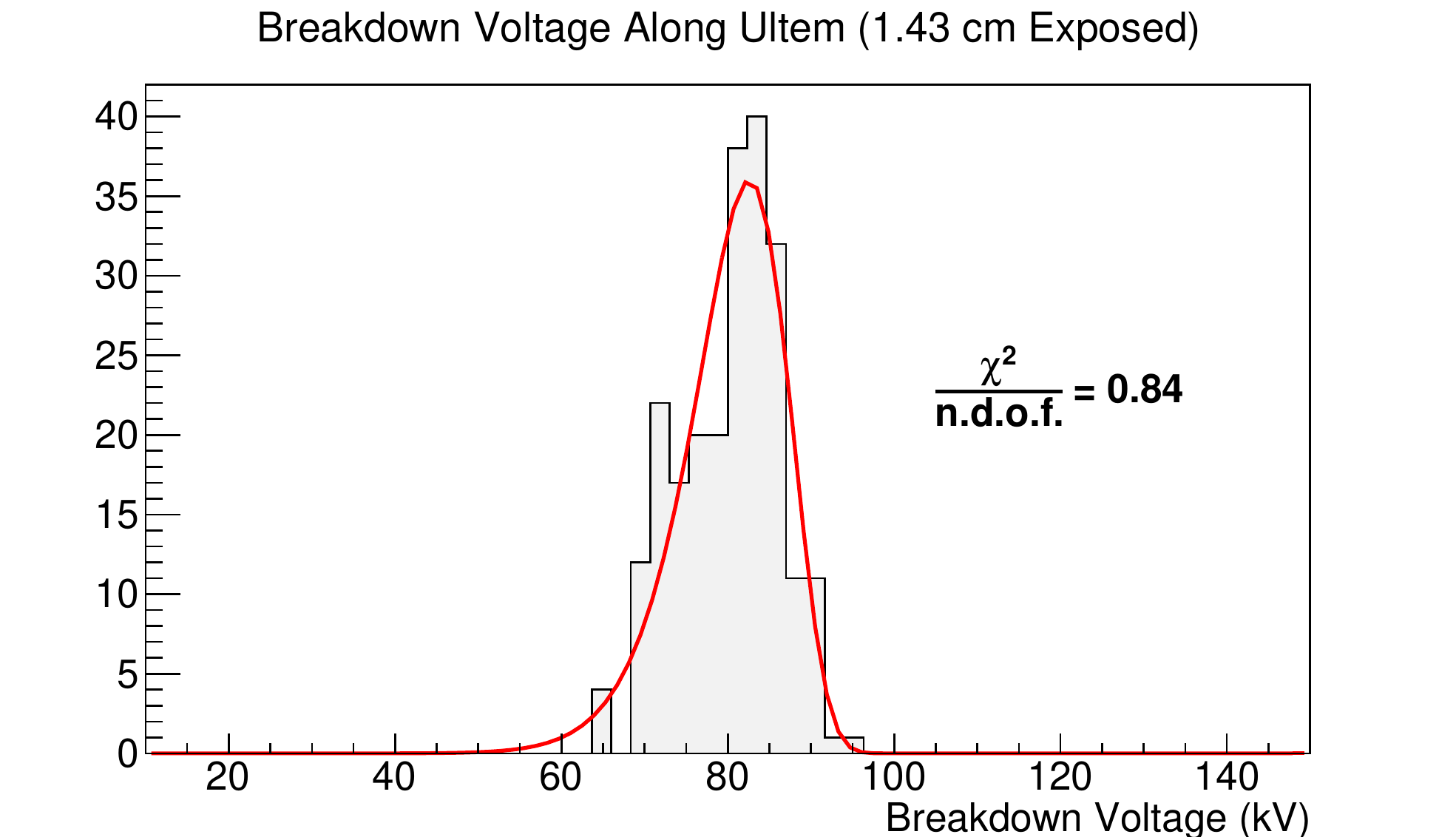}}
\caption{The sequential breakdown voltages for the Ultem test piece with 1.43~cm exposed is shown in (a), and a Weibull functional fit to the resulting distribution is shown in (b).}
\label{fig:ultemExample}
\end{figure}

We evaluated the 11 materials listed in Table~\ref{tab:shrinkage}.  
Tested ring elevations varied for each sample, but at least three exposed insulator lengths were evaluated between 3.5-12 cm for each undamaged smooth sample.  
For a given test piece and length of exposed insulator above the grounded ring, the number of trips tested ranged from 42-297 depending on the available time before refilling the argon.  The voltage values of these trips were plotted sequentially to check for stability, and then histogrammed and fit with a Weibull function~\cite{ref:weibull1951} as shown for example in Figure~\ref{fig:ultemExample}.  The Weibull function is described elsewhere~\cite{ref:gauster} as being an appropriate function to describe the variability of breakdown voltages. The mean and the square root of the variance were taken from the fit to compare breakdown voltages versus lengths between materials.

\begin{table}
\centering
\footnotesize
\begin{tabular}{@{} l l l l r r @{}}
\hline \hline{}
                & Thermal           & Chemical                    & Chemical                & Dielectric  & Dielectric \\
   Material     &Contraction        & Name                        & Abbreviation            &  Constant   & Strength   \\
                &(\%)&&&& (kV/mm)\\ \hline
\multirow{2}{*}{Delrin}& \multirow{2}{*}{1.2 $\pm$ 0.1}   & \multirow{2}{*}{Polyoxymethylene-homopolymer}  & Acetal-Homopolymer   &  \multirow{2}{*}{2.7}	     & \multirow{2}{*}{20}     \\
	        &    &   & POM   &       &      \\
 Polypropylene	& 1.1 $\pm$ 0.09  & Polypropylene	          & PP	                     &  2.2-2.6	     & 30-40  \\
 Noryl	        & 0.5 $\pm$ 0.08  & Polyphenyleneoxide+Polystyrene& PPO + PS	             &  2.7	     & 16-20  \\
 Polycarbonate	& 0.6 $\pm$ 0.09  & Polycarbonate          	  & PC	                     &  2.9	     & 15-67  \\
 Polystyrene	& 0.5 $\pm$ 0.06  & Polystyrene	                  & PS	                     &  2.4-3.1	     & 20     \\
 UHMW PE	& 1.8 $\pm$ 0.3   & Polyethylene-UHMW	          & UHMW PE	             &  2.3	     & 28     \\
 FR4	        & 0.2 $\pm$ 0.07  & Fiberglass-epoxy	          & G10-FR4	             &  4.8	     & 20     \\
 PTFE	        & 0.9 $\pm$ 0.09  & Polytetrafluoroethylene	  & PTFE	             &  2.0-2.1	     & 50-170 \\
 Ultem      	& 0.4 $\pm$ 0.07  & Polyetherimide	          & PEI	                     &  3.1	     & 30     \\
 PBT	        & 0.9 $\pm$ 0.08  & Polyethtylene Terephtalate	  & PBT	                     &  3.2	     & 20     \\
 Polyester	& 0.9 $\pm$ 0.1   & Polybutylene Terephtalate	  & PET, PETB	             &   3	     & 17     \\
\hline \hline
\end{tabular}
\caption{The thermal contraction, dielectric constants, and dielectric strengths for the materials under test.  The thermal contraction was measured between room temperature and 78~K.  The Delrin dielectric constant was referenced from~\cite{ref:dupontDelrin}, the FR4 values from~\cite{ref:wikiFR4}, all others are from~\cite{ref:professionalPlastics}.}
\label{tab:shrinkage}
\end{table}
\FloatBarrier

\FloatBarrier
\section{Results}

\subsection{Cold Shrinkage Measurements}
Before electrically testing the insulators, their thermal contraction was measured.  The diameter of 5.1~cm rod stock of each material was measured at room temperature and
at liquid nitrogen temperature.  The technique was to submerge each piece in liquid nitrogen until there was no visible boiling.  Each test piece was then removed and measured at three common reference lengths evaluating the minimum and maximum diameters.  The values are reported in Table~\ref{tab:shrinkage}.  These values are useful for understanding the capabilities and limits of components of LArTPCs, and planning for dimensional change at cryogenic temperatures.

\FloatBarrier

\subsection{Electrical Insulator Test Results}

The insulator test pieces can be classified into three groups:  those that catastrophically fractured upon submersion in liquid argon, those that broke after a number of successful dielectric breakdown tests, and those that underwent the full suite of testing.  Here, a successful breakdown test is defined as a dielectric breakdown near the surface of the insulator that trips the power supply.  For some materials, additional testing of surface profiles was performed.  This testing is discussed in Section~\ref{sec:grooves}.

\subsubsection{Group I}

Polystyrene fractured upon submersion in liquid argon. 
The Noryl test piece made a popping noise as it was lowered into the liquid argon.  The piece was then removed for inspection and burst as it warmed. 
The contraction of the solid stock of both materials had been measured without sign of damage.  It appears that introducing the non-contracting center conductor led to too much strain for the material when exposed to liquid argon temperatures.  These materials are problematic for use in high voltage feedthroughs.

\subsubsection{Group II}
The next group of materials was able to undergo a number of successful breakdown measurements, but eventually suffered mechanical damage preventing further testing.  This group included polyester, PBT, and PTFE.  
Polyester and PBT both suffered catastrophic mechanical failures while testing with 11.43~cm of material exposed.  Polyester failed after 50 breakdown events, and PBT lasted through 100.

\begin{figure}[hbt]
  \centering
  \includegraphics[width=0.45\textwidth]{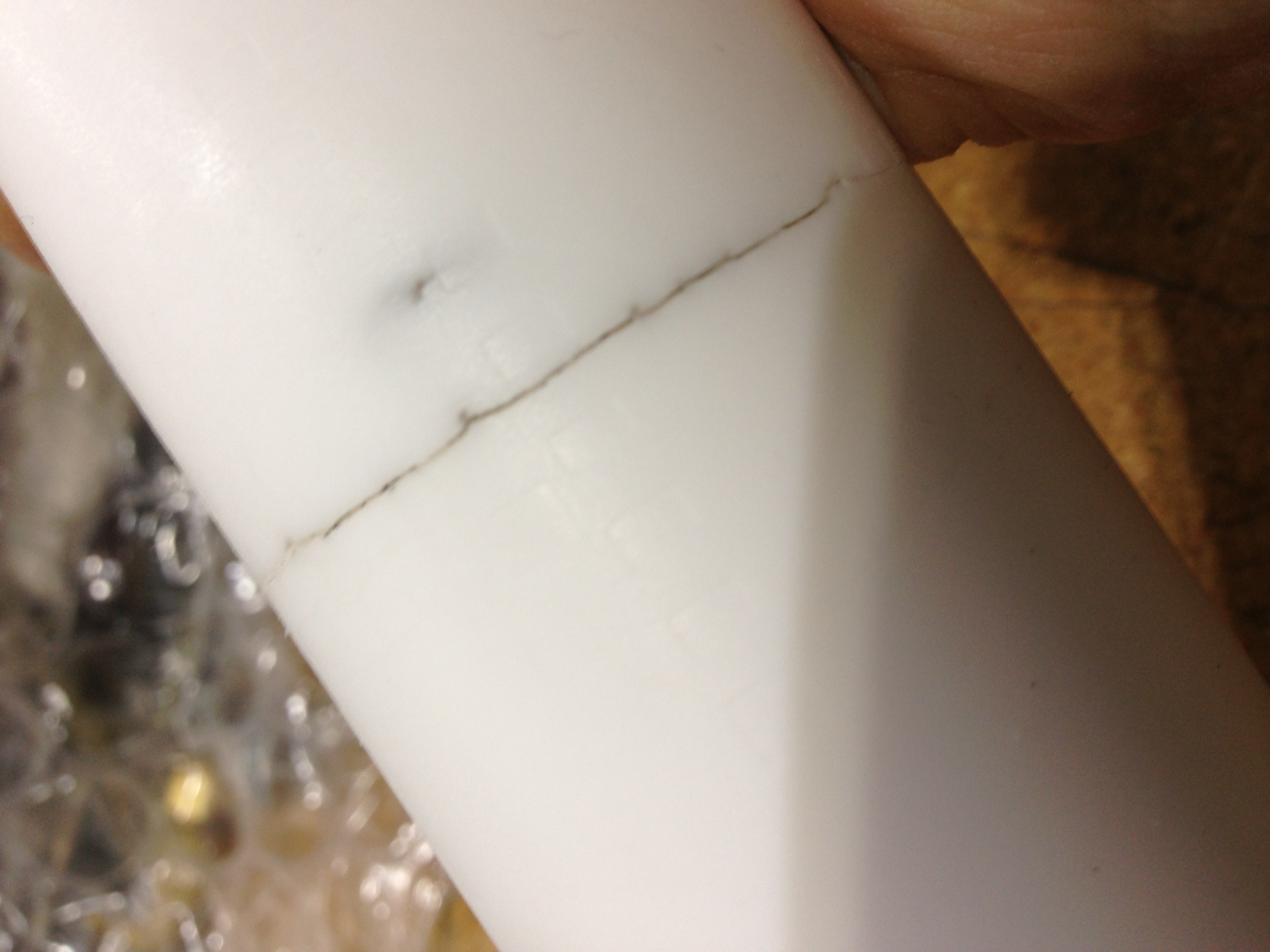}
\caption{The PTFE test piece after evaluation.  }
\label{fig:teflonPhotos}
\end{figure}

PTFE is often used in cryogenic devices, and initially showed much promise in this study
with average breakdown values among the highest evaluated.  However, while testing with 11.43~cm length exposed, the breakdown voltage suddenly dropped after about 17 breakdowns.  Upon inspection, it was found that a pinhole had burned through the bulk of the material and a major crack developed around the diameter nearly severing the piece as seen in Figure~\ref{fig:teflonPhotos}.  This was the only material that broke through the bulk near the base of the center conductor.  The peak field at the bottom corner of the center conductor was $\sim$400~kV/cm which is in the range of the published dielectric strength of the material.

\FloatBarrier

\subsubsection{Group III}
The remaining materials survived mechanically during testing.  All had stable breakdown voltages in time showing no degradation after hundreds of discharges.  

\begin{figure}[hbt]
  \centering
   \subfloat[]{\includegraphics[width=0.50\textwidth]{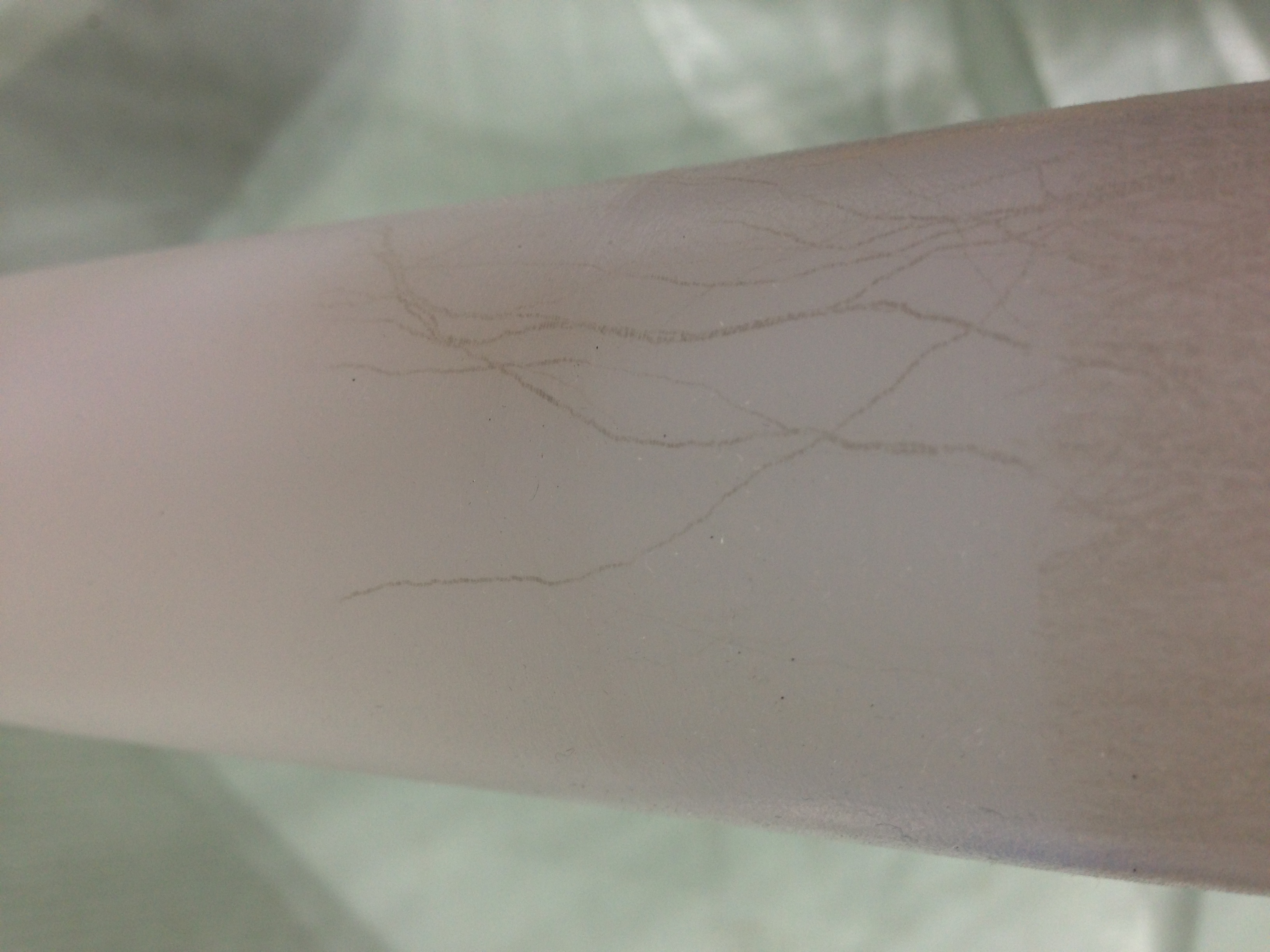}}
   \quad
   \subfloat[]{\includegraphics[width=0.44\textwidth]{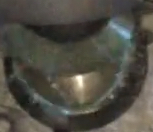}}
\caption{Traces along the UHMW PE test piece are shown in (a).  In the right of (a), the traces are more dense due to many more electric breakdowns being evaluated at that elevation.  In (b), a glowing path from breakdown along the surface of the FR4 test piece is shown.  The glowing trace goes on a path from the ring along the right side of the piece up the surface and then radially inward to the center conductor.}
\label{fig:traces}
\end{figure}

Polypropylene and UHMW PE had the highest average breakdown voltages.
UHMW PE exhibits light brown traces about 0.2-0.5~mm wide and of uniform contrast along their length on the surface presumably where the spark traveled.  UHMW PE displays this tracing, seen in Figure~\ref{fig:traces}, more clearly than any of the other materials.  These traces show that the breakdowns follow the surface closely and we found that new traces formed as there were more breakdowns showing that breakdowns do not preferentially follow existing traces.

\subsubsection{Analysis of Group III}
\begin{figure}[hbt]
   \centering
\includegraphics[width=0.8\textwidth]{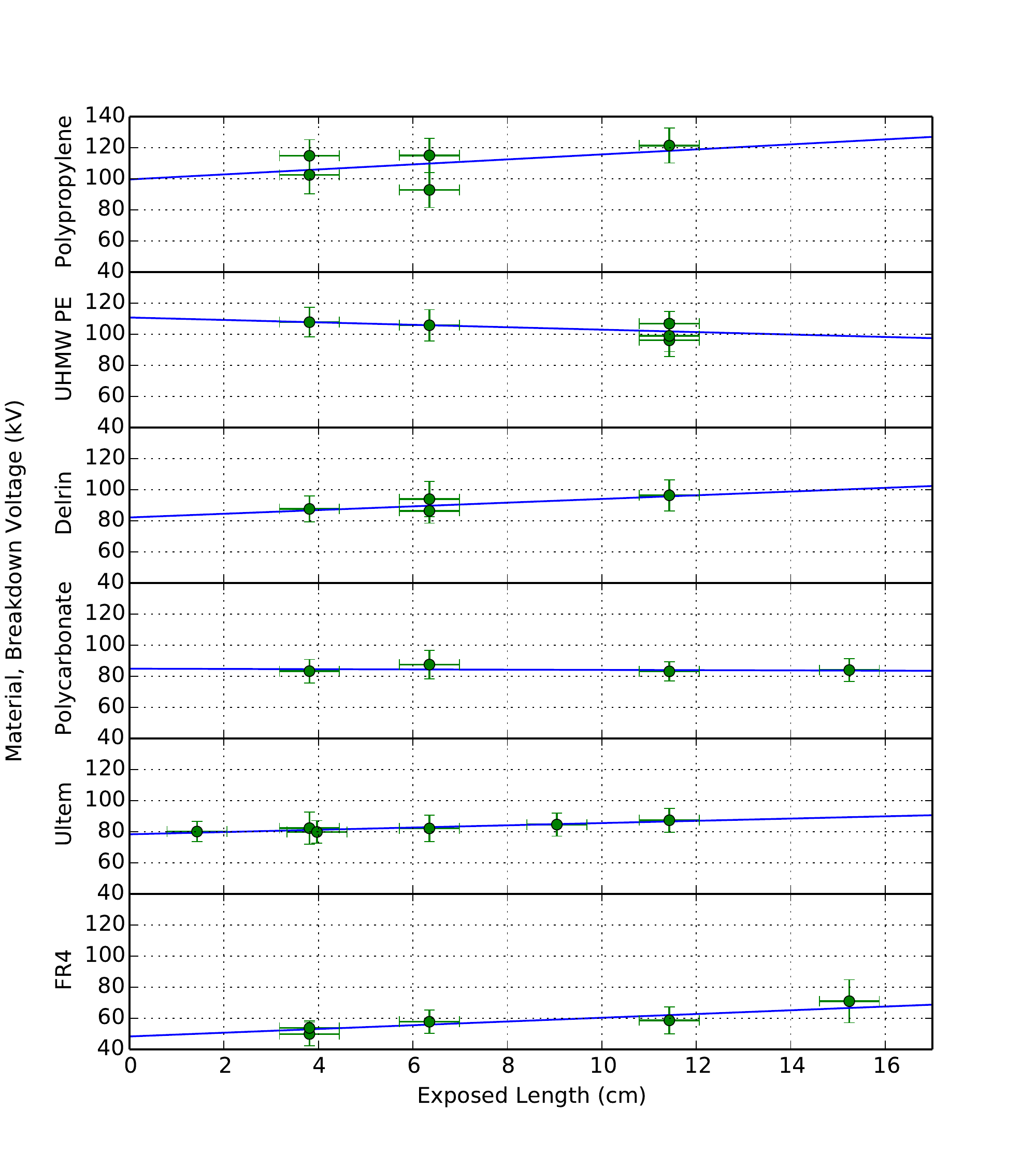}
\caption{Mean breakdown voltages as a function of exposed insulator length for the materials that survived testing.  Figure~\protect\ref{fig:slopes} gives the values of the slopes, ($\Delta$V)/($\Delta$exposed length).  Some configurations were repeated at different times during the run of the experiment resulting in multiple data points for a given material's exposed length.}
\label{fig:group3}
\end{figure}

\begin{figure}[h]
  \centering
  \includegraphics[width=0.70\textwidth]{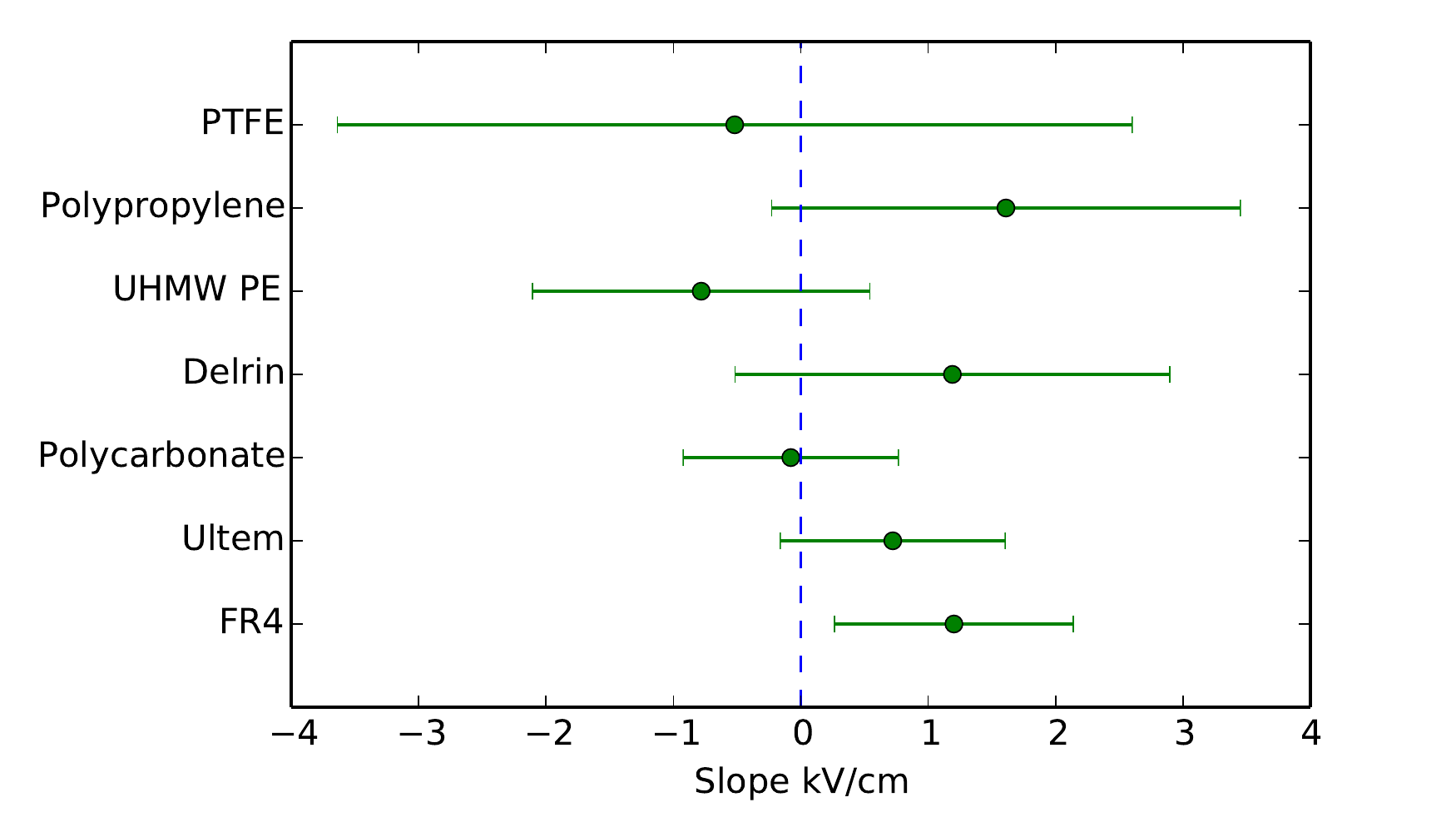}
\caption{Slopes from the fit to the breakdown voltage versus exposed insulator length data.}
\label{fig:slopes}
\end{figure}

We now turn to the measurements made on the materials of Group III:  those that survived the mechanical tests and hundreds of discharges.
The average breakdown voltages from the Weibull fit for the Group III materials are plotted versus exposed insulator length in Figure~\ref{fig:group3}.  A linear fit of the resulting averages is also shown with the slopes reported in Figure~\ref{fig:slopes}.  As can be seen from the fits, there is only a very weak dependence, if any, of the breakdown voltage on exposed length.

\begin{figure}[hbt]
   \centering
\includegraphics[width=0.8\textwidth,trim={0cm 0cm 7cm 0cm},clip]{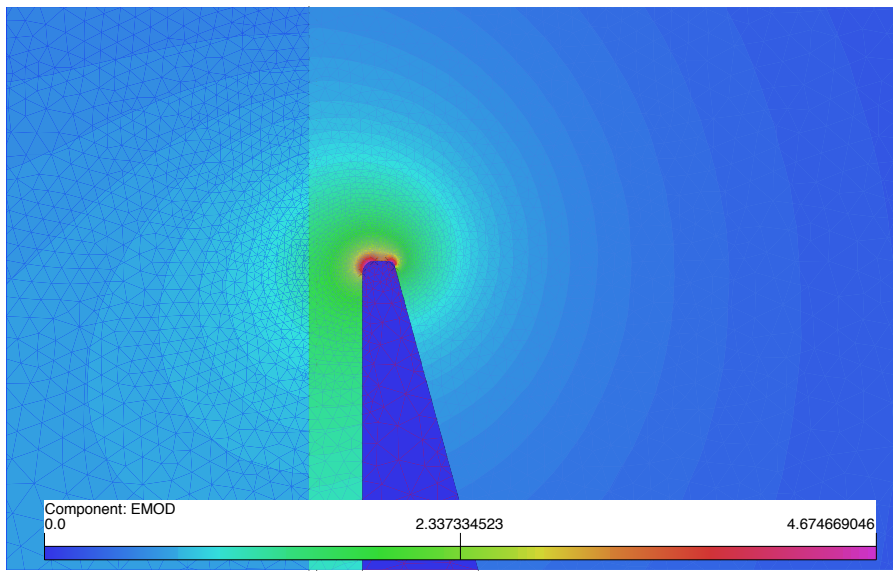}
\caption{The magnitude of the electric field near the ring tip from the FEA of UHMW PE with a 6~cm exposed length.  Here, 1~V is applied; the field (V/cm) from a desired voltage can be scaled.  The insulator is on the left; the center conductor is on the left beyond the frame of the image.}
\label{fig:ringFEA}
\end{figure}

The material's dielectric constant was the primary driver of the difference in the field at the ring for a given voltage.  An FEA (Finite Element Analysis) of the setup was performed for different materials with a 0.06~cm gap between the ring and the insulator surface; an example FEA near the ring is shown in Figure~\ref{fig:ringFEA}.  For a 6~cm exposed insulator on UHMW PE, the peak field at 100~kV was $470 \pm 10$~kV/cm.  Shrinking the insulator diameter to the minimum and maximum of the thermal contractions led to a change of only 1.2\% and 1.6\% in peak electric field when keeping the dielectric constant fixed.


\clearpage

\FloatBarrier

\subsection{Insulators with Grooved Profiles}\label{sec:grooves}

In addition to material and exposed insulator length, the surface profile of an insulator is also a parameter of interest.  Ridged or grooved surface profiles are commonly used in high voltage bushings.  For example, the often seen ceramic bushings used in outdoor power distribution are designed to prevent flash-over by increasing the creepage path and keeping it dry~\cite{ref:alston}.  Grooves have also been employed in high voltage feedthroughs in liquid argon \cite{ref:icarus}.  In this section, we present a study of the performance of profiles in liquid argon.

\begin{figure}[h]
  \centering
   \includegraphics[width=0.6\textwidth]{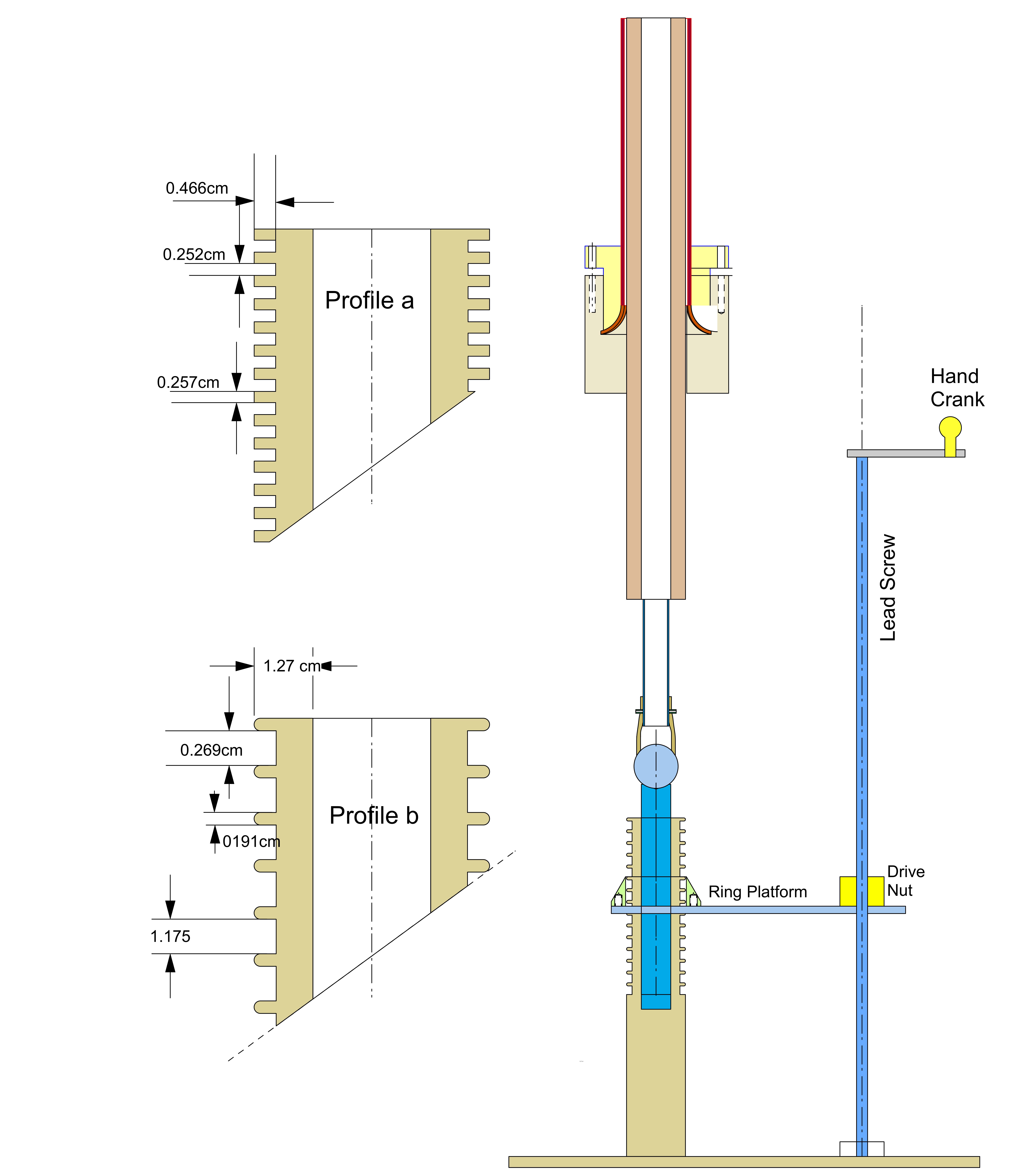}
   \caption{The two profiles used in testing, and the test apparatus.}
\label{fig:profiles}
\end{figure}

In UHMW PE, the profiles shown in Figure~\ref{fig:profiles}(a) and (b) were prepared to compare ridge shape and spacing.  In FR4, and polycarbonate, only the profile shown in Figure~\ref{fig:profiles}(b) was evaluated.

\begin{figure}[hbt]
  \centering
  \subfloat[]{\includegraphics[width=0.5\textwidth]{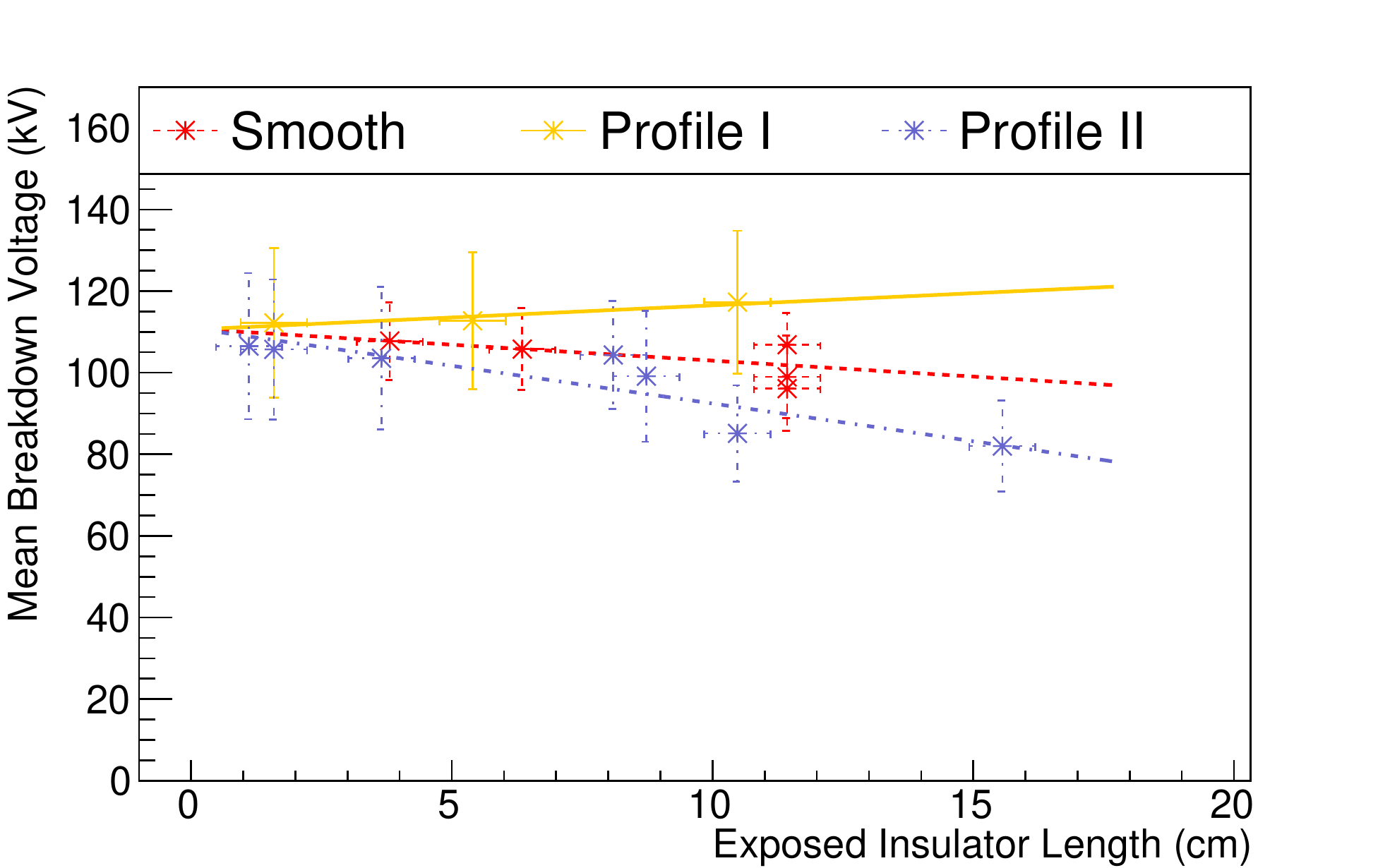}}
  \subfloat[]{\includegraphics[width=0.50\textwidth,trim={1cm 0cm 1cm 0cm},clip]{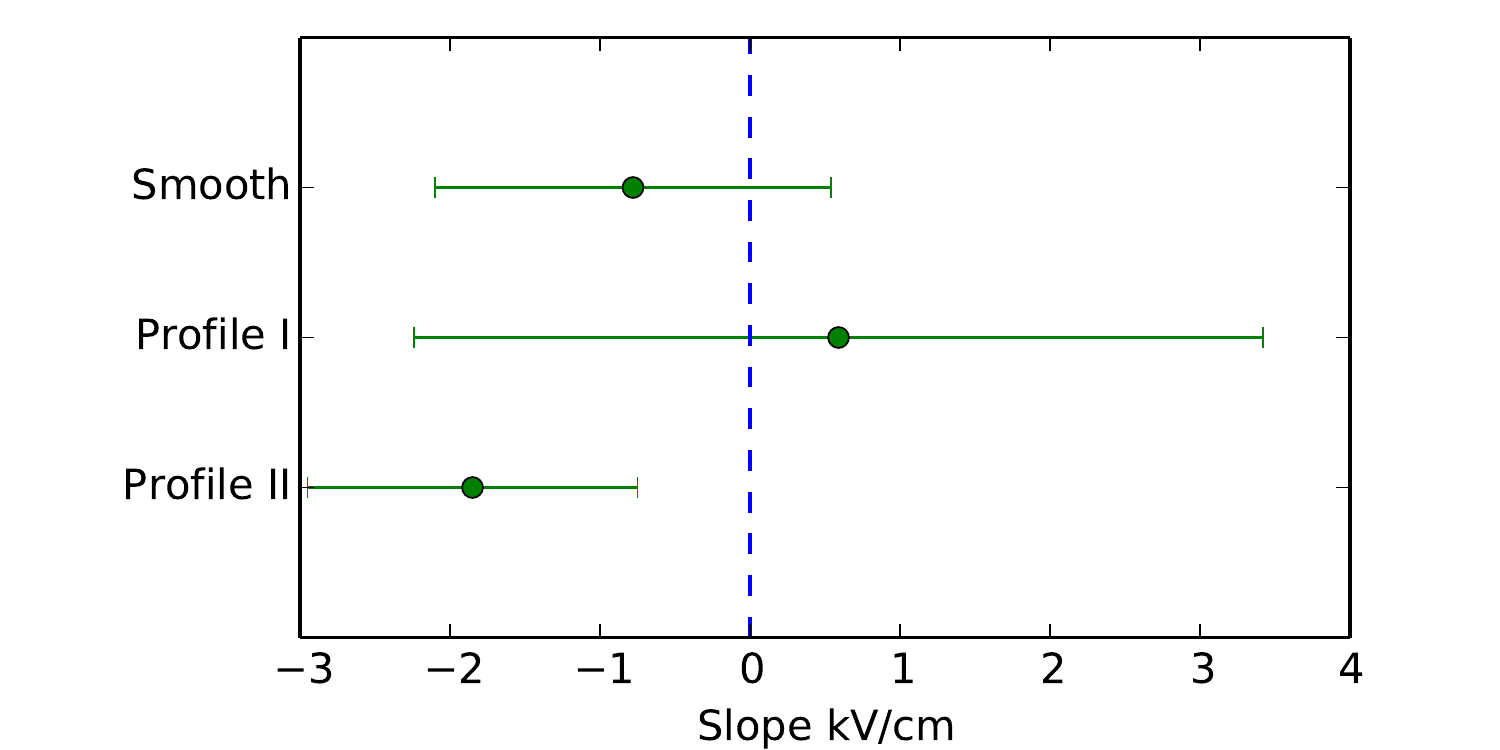}}
  \caption{Summary plots for the mean breakdown voltage values as a function of exposed insulator length for the UHMW PE samples. }
\label{fig:voltages_groove_all}
\end{figure}


The results from the grooved samples show at most a small effect on breakdown voltage.  The results for the smooth and grooved UHMW PE are shown in Figure~\ref{fig:voltages_groove_all}. 
It is worth noting that the traces were observed to follow the surface of the grooves on the UHMW PE samples.

The remaining grooved materials all suffered damage of some sort.  The grooved polycarbonate test piece developed damaged ridges during testing.  Only one distance at 4.13~cm was evaluated and it is difficult to determine if the ridges were made weaker during the initial installation of the ring, or if the ridges were only damaged from electrical discharge.  The grooved FR4 piece also had only one exposed length at 6.67~cm evaluated due to damage.  Here, however, the ridges were not damaged, but rather a crack burned through from the center conductor outward along fibers of the FR4.

\section{Discussion}\label{sec:discussion}


We have evaluated insulator performance by studying dielectric breakdown along the surface of insulators in liquid argon in an open cryostat.  In our setup, a high voltage conductor was encased in the insulator under test.
We have studied a variety of candidate materials and compared their performance.  We have also evaluated the effect of exposed insulator length on breakdown voltage.

\subsection{Dependence of Breakdown Voltage on Material}

\begin{figure}[hbt]
  \centering
  \includegraphics[width=0.70\textwidth]{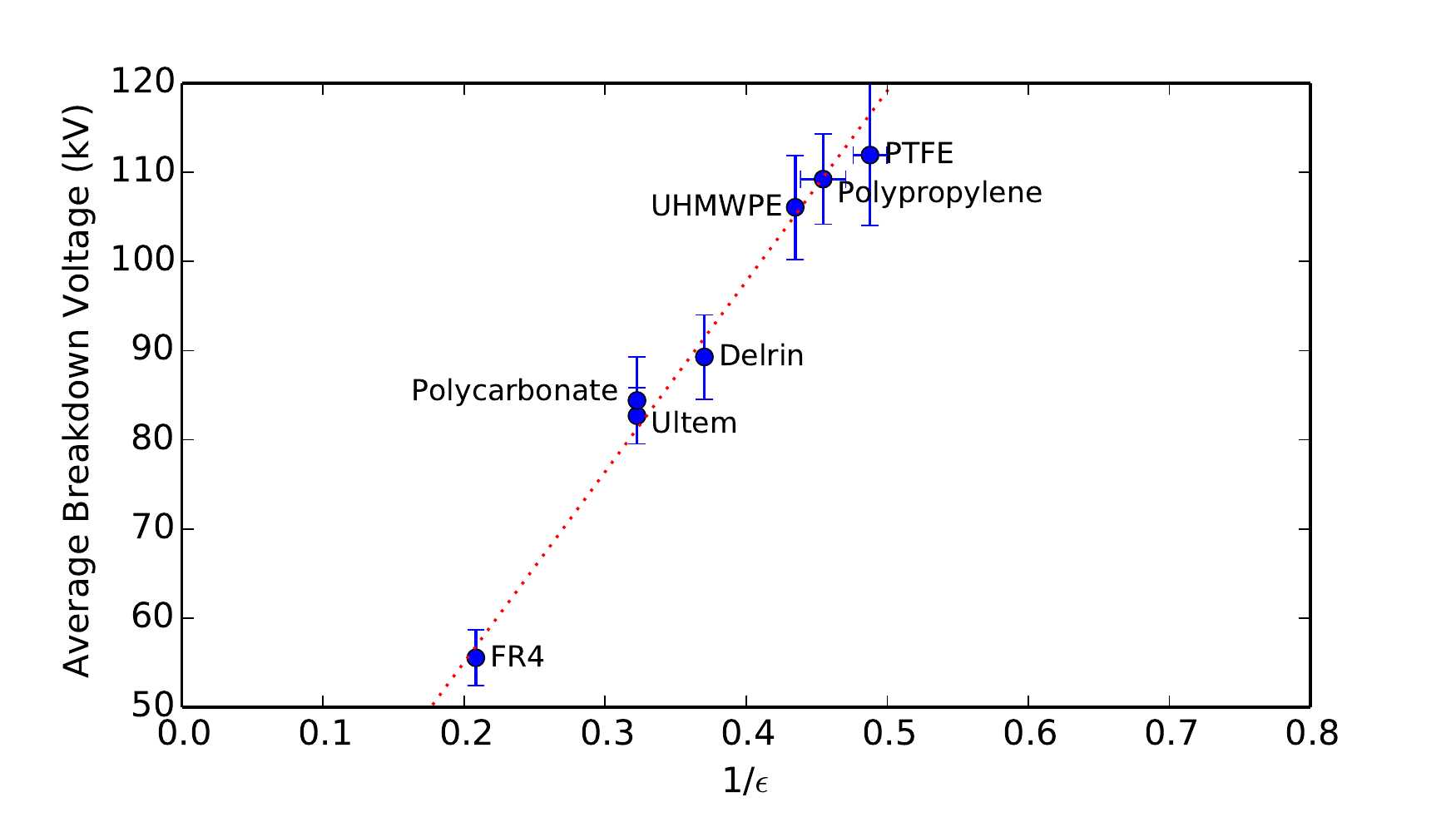}
\caption{The breakdown voltage for several materials at an exposed insulator length of 6~cm versus the inverse permittivity.}
\label{fig:averageBDPermittivity}
\end{figure}

The average breakdown voltage for a common exposed insulator length of 6~cm is shown in Figure~\ref{fig:averageBDPermittivity} versus the inverse of the materials' permittivities.  Materials with high permittivity increase the field near the knife edge on the ring.
We find that the material dependence of breakdown voltage can be entirely understood as a consequence of the sample's permittivity and the resulting electric field near the insulator surface, and does not display any other material effects.

\subsection{Dependence of Breakdown Voltage on Exposed Insulator Length}

From our testing, we found little, if any, dependence on breakdown voltage versus exposed insulator length as shown in Figure~\ref{fig:slopes}.  This result differs from
broad experience at room temperature in the high voltage community
that generally relates an increase in creepage path to an increase in breakdown voltage.

In an attempt to understand breakdown behavior in liquid argon, we turned to the literature on breakdown in oil.  
A paper by Meek~\cite{ref:meek} is one of the first to tackle breakdown in dense media, and provides a narrative for the breakdown chain of events.  This paper states, ``The breakdown of a uniform field is considered to occur by the transition of an electron avalanche proceeding from cathode to anode into a self-propagating streamer, which develops from anode to cathode to form a conducting filament between the electrodes.''  This model of an initializing avalanche followed by streamer propagation is still accepted today.

\begin{figure}[hbt]
  \centering
  \includegraphics[width=0.6\textwidth]{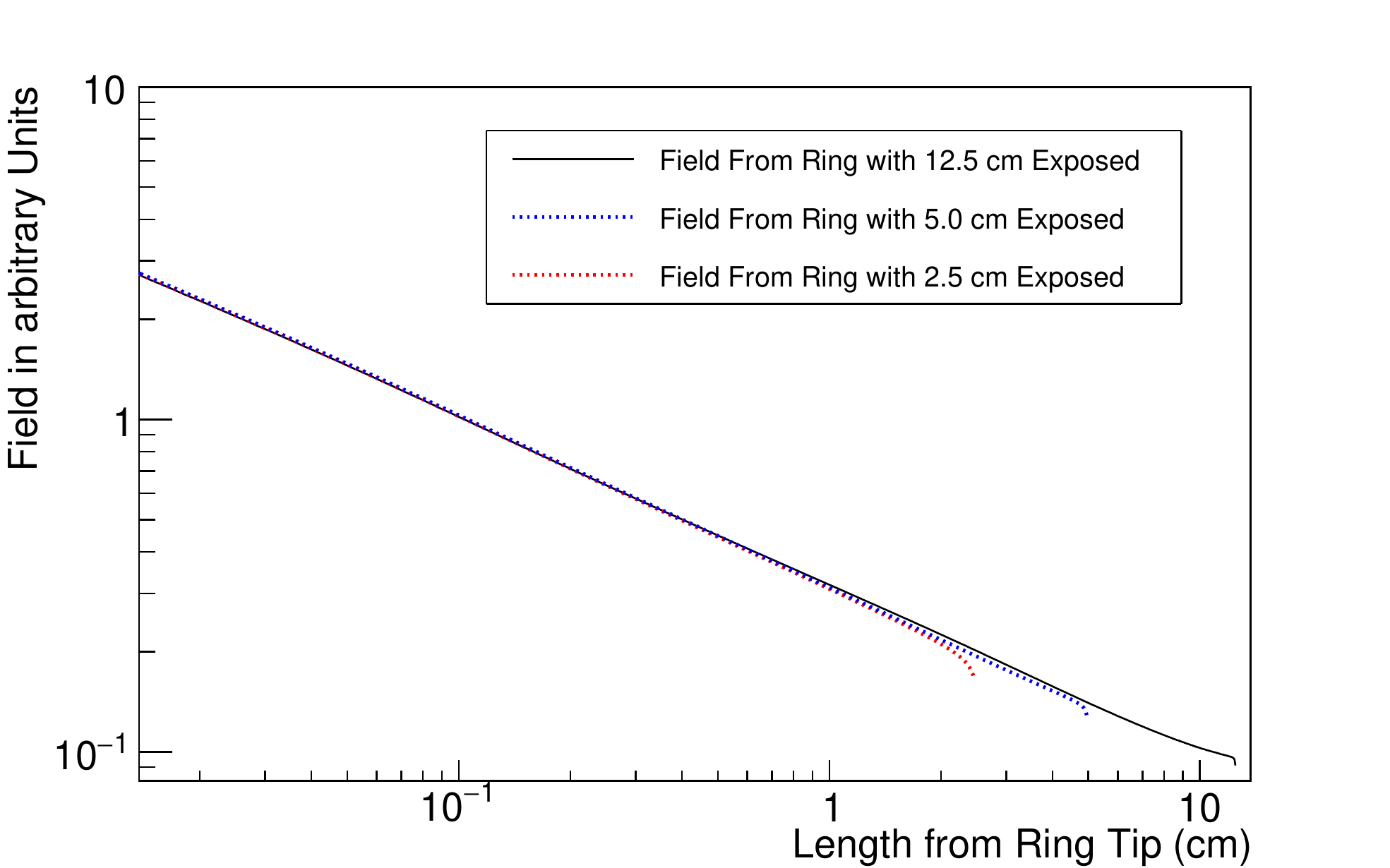}
\caption{The electric field magnitude along the vertical direction from the ring tip for three exposed insulator lengths.  The dominant field near the ring tip is independent of exposed insulator length.}
\label{fig:ezfield}
\end{figure}

In our setup, there is a concentrated electric field by design near the sharp feature at the knife edge of the ring.  This field is roughly independent of ring depth as shown in Figure~\ref{fig:ezfield} where different exposed lengths yield a similar electric field from the ring tip vertically along the insulator length.  The sharp field enables the initialization of a streamer.  

Once started, the streamer propagates toward the cathode.
The streamer grows by ionization from electrons and photons; therefore it needs energy to propagate.  In Reference~\cite{ref:fridman}, a mechanism is described, ``The cathode-directed streamer starts near the anode. It looks like and
operates as a thin conductive needle growing from the anode. The electric
field at the tip of the `anode needle' is very high, which stimulates the fast
streamer propagation in the direction of the cathode.''

In our test setup, just as the electric field on the edge of the ring created a field strong enough to ionize the argon, the highly localized charge in the streamer head concentrates the electric field from the center conductor supporting further ionization for streamer growth.  The field from the center conductor is independent of elevation and is the dominating field along the insulator.  This field grows the streamer without any need for additional fields along the insulator surface.  Hence, the voltage required for breakdown is independent of exposed insulator length.

Our results of breakdown voltages as a function of exposed insulator length are shown in Figure~\ref{fig:group3} and the resulting slopes from the fits are summarized in Figure~\ref{fig:slopes}.  We see no significant material dependence, and further, the central values are
consistent with zero as one would expect from the model described.

\subsection{Dependence of Breakdown Voltage on Insulator Profile}\label{sec:discussionGrooves}

A number of arguments have been made to motivate the addition of grooves to high voltage components in liquid argon.  Path length arguments are popular,
as is the suggestion that creating a surface perpendicular or in opposition to the electric field inhibits the travel of charge.  We observed at most a modest effect from the addition of grooves to our test pieces.  

\begin{figure}[hbt]
  \centering
  \subfloat[]{\includegraphics[width=0.50\textwidth]{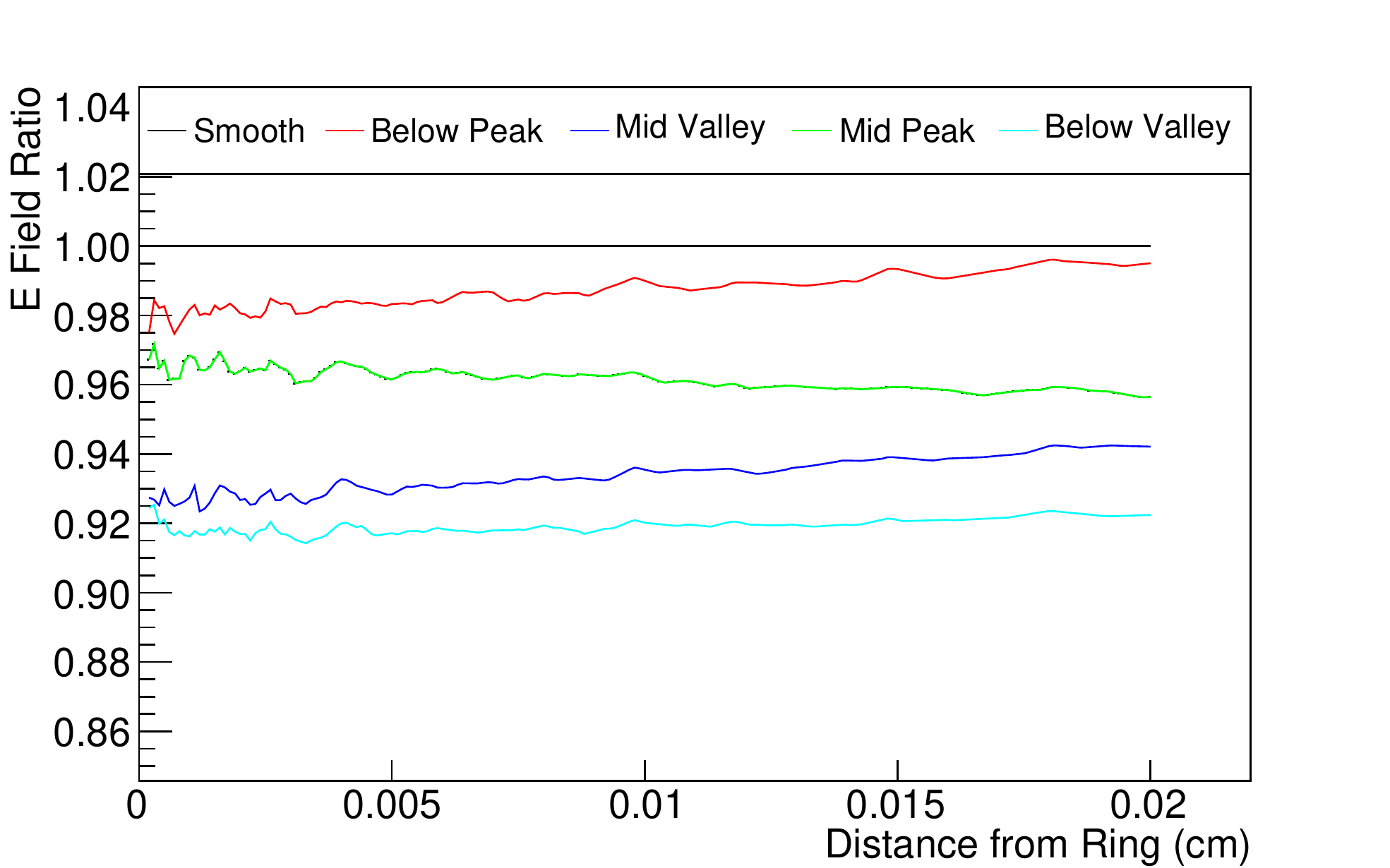}}
  \subfloat[]{\includegraphics[width=0.50\textwidth]{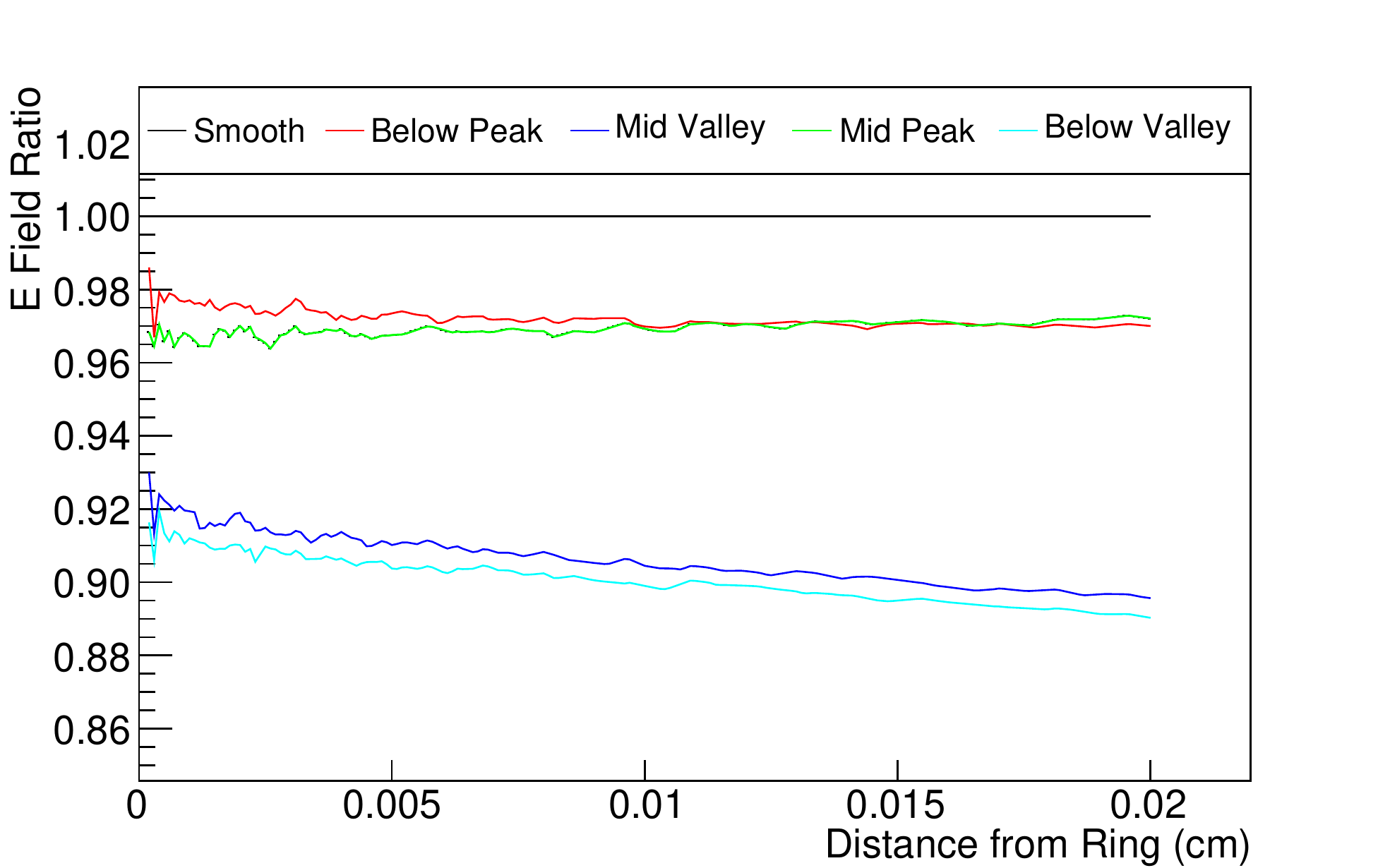}}
\caption{Ratio of the electric field near the ring for different exposed insulator lengths with respect to the groove profile.  Figure (a) is the ratio of the magnitude of the electric field in a line directed radially toward the sample from the ring tip.  Figure (b) is the ratio of the magnitude of the electric field in a vertical line from the ring tip.}
\label{fig:eFieldsRatioGrooves}
\end{figure}

At least part of the efficacy of grooves in our tests is due to the reduction in field by replacing plastic with argon near the grounded ring.  During testing, we did not know the precise location of the ring tip relative to the grooves.
FEAs of the field at a common cathode voltage for Profile II grooves were performed for different ring heights relative to the groove.  A reduction in field can be seen in Figure~\ref{fig:eFieldsRatioGrooves} where the ratio of the magnitude of the field for a grooved sample to that of a smooth sample is shown for a common input voltage.
Similar results were found for polycarbonate and FR4.  The largest effect is seen when the ring is between the middle of a valley of a groove and the top edge of a peak.  Here, the ratio of the field at a common voltage of the smooth insulator, $E_s$, to the field of the grooved insulator, $E_G$, was nearly 0.90 for UHMW PE, 0.84 for polycarbonate, and 0.75 for FR4.  These values along with the observed breakdown voltages are summarized in Table~\ref{tab:grooves}.
One can scale the modeled peak electric fields by the observed breakdown voltages, $V_S$ and $V_G$ for smooth and grooved samples respectively, to obtain the field at breakdown.  If one assumes there is a similar electric field at breakdown between the geometries, then $V_SE_S \simeq V_GE_G$.  If $(V_S/V_G) \leq (E_G/E_S)$, the increase in breakdown voltage of the grooved samples can be accounted for by the reduction in field due to the change in the permitivity by removing the insulator material.  For polycarbonate and FR4, the ratios $E_G/E_S$ are in rough agreement with the observed average breakdown voltages ratio $V_S/V_G$.

\begin{table}
\centering
\small
\begin{tabular}{@{} l c c c r r @{}}
\hline \hline
\multirow{2}{*}{Sample}& Grooved Exposed& Grooved                         & Smooth            &  FEA      & Measured  \\
                       & Length (cm)  &$<V_{BD}>$                        &$<V_{BD}>$         &$E_G/E_S$     &$V_S/V_G$ \\\hline
FR4                    &  6.67        & $78.1\pm14.3$                   & 55.9              & 0.75         & 0.72 \\
Polycarbonate          &  4.13        & $98.4\pm13.3$                   & 84.6              & 0.84         & 0.86 \\
UHMW PE                &  1.11---15.56 & $106.5\pm17.9$ --- $82.0\pm11.1$ & 109.9---98.6       & 0.90         & 1---1.2 \\
\hline \hline
\end{tabular}
\caption{The average breakdown voltage, $V_{BD}$, for smooth and Profile II grooved samples, and their resulting ratio, $V_S/V_G$.  The smooth value is from the fitted line to the breakdown voltages as a function of exposed length.  Also included is the minimum ratio of the electric fields vertically from the ring tip for the samples as computed by an FEA.  Here the ring tip is just below a valley and this is where the maximum effect on the electric field is seen.}
\label{tab:grooves}
\end{table}

The UHMW PE Profile II sample had a  measured $V_S/V_G>1$ in contrast to the other samples, including the UHMW PE Profile I sample, and the FEA model.  The precise effect of grooves on performance is not clear, however, they do not appear to have a large effect on breakdown voltage in this study as shown in Figure~\ref{fig:voltages_groove_all}(b) where the slopes and errors are reported for the grooved and smooth UHMW PE.  The values are again consistent with zero.


\section{Conclusion}

We have performed a study of electrical breakdown in liquid argon along the surface of insulators surrounding a conductor with regard to breakdown path length, insulator material, and insulator surface profile.  The measurements were performed in an open-to-air cryostat initially filled with commercial grade argon, and the breakdown was initiated at the anode.  We identified some materials that failed mechanically in our test setup.  We found the breakdown voltage does not depend upon the length of the exposed insulator.  Material-specific effects could be explained as due to their permittivities.

The evaluation of surface profiles point to at most a modest effect on breakdown voltage by adding grooves.
Two grooved materials failed mechanically after one exposed length test.  Before failing, the breakdown voltage recorded was lower when compared to a smooth sample.  However, this change in breakdown voltage could be accounted for by the change in the peak electric field resulting from the grooved material geometry.
A limited study of exposed length of grooved UHMW PE pointed to little, if any, effect on breakdown voltage with all of the average voltage gains with length being
consistent with zero.

A practical consequence of our observations (in impure liquid argon) is that for insulator tubes surrounding high voltage electrodes, the peak surface electric field is the relevant value.  Thus, the permittivity of the materials, provided they can survive mechanically, is an important parameter of concern.  Adjusting the radial dimension will likely also improve performance.  Longer insulators provide little, if any, improvement in high voltage standoff.  Grooving the insulator surface similarly seems to have little effect beyond reducing the local field.  Improvements must focus on preventing avalanche initiation, or preventing the avalanche to escape to where it can grow to be a longer streamer.

\section{Acknowledgements}
The authors wish to thank Brian Rebel and Stephen Pordes for their support, editorial contributions, and scientific discussions, and Jim Walton and Alan Hahn for their technical assistance.

Fermilab is operated by Fermi Research Alliance, LLC under Contract No. De-AC02-07CH11359 with the United States Department of Energy.

\bibliographystyle{utphys}
\bibliography{references}

\end{document}